\documentstyle[aps,preprint,epsf]{revtex}

\begin{document}
\title{Can Hall drag be observed in Coulomb coupled quantum wells
in a magnetic field?} 

\author{Ben Yu-Kuang Hu\\
\smallskip
Mikroelektronik Centret,
Bygning 345 \o st, Danmarks Tekniske Universitet,\\
DK-2800 Lyngby, Denmark
}

\date{\today}
\maketitle

%%%%%%%%%%%%%%%% ABSTRACT %%%%%%%%%%%%%%%%%%%%%%%%%%%%%%
\begin{abstract}

We study the transresistivity $\tensor\rho_{21}$ (or equivalently, 
the drag rate) of two Coulomb-coupled quantum wells in the 
presence of a perpendicular magnetic field, 
using semi-classical transport theory. Elementary arguments seem to preclude
any possibility of observation of ``Hall drag'' (i.e., 
a non-zero off-diagonal component in $\tensor\rho_{21}$).
We show that these arguments are specious, and in
fact Hall drag can be observed at sufficiently high temperatures 
when the {\sl intra}layer transport time $\tau$ has 
significant energy-dependence around the Fermi energy $\varepsilon_F$.
The ratio of the Hall to longitudinal transresistivities 
goes as $T^2 B s$, where $T$ is the temperature, $B$ is the 
magnetic field, and $s = [\partial\tau/
\partial\varepsilon] (\varepsilon_F)$.
\end{abstract}

%\pacs{}

%%%%%%%%%%%%%%%% INTRODUCTION %%%%%%%%%%%%%%%%%%%%%%%%%%%%%%

\newpage

\section{Introduction}

When two quantum wells are placed sufficiently close together (but
far enough apart so inter-well tunneling is negligible), 
Pogrebinskii and Price\cite{originate} predicted that 
the interlayer electron--electron ($e$--$e$) interactions would be enough
so that a drift of carriers in one layer causes a discernable
drag in the other.  Such a ``Coulomb drag'' effect has been 
measured\cite{exper} using in a set-up shown schematically 
Fig.\ 1, and its observation has prompted a barrage of 
theoretical studies\cite{theory,jauh93,kame95,flen95a,flen95b,bons96}.

Experimentalists are now extending their studies to Coulomb drag in 
systems with a magnetic field $B$ perpendicular to the layers\cite{expmag}. 
Present efforts are mainly focussed
on the high $B$-field regime, when Landau levels are fully resolved,
and interesting effects from the quantization have been
predicted\cite{bons96}. In this regime, the drag electric field 
response ${\bf E}_2$ was shown using the Kubo formalism to be
parallel to the driving current ${\bf J}_1$ (to lowest 
non-vanishing order in the interlayer interaction and $1/B$). 
The question naturally arises: are there circumstances when  
one can find ${\bf E}_2$ with a component {\sl perpendicular} 
to ${\bf J}_1$; i.e., Hall drag?  
And if so, what does this tell us about the system?

In this paper, we first present a seemingly plausible explanation
for why, at the level of the Born approximation, Hall drag
{\sl cannot} exist.  We then reveal the flaw behind the  
explanation and show that Hall drag is possible in principle. 
Using a low temperature ($T$) expansion, we show that the 
Hall transresistivity should go as $T^4$, and that the  
magnitude of the Hall drag gives information on the 
energy-dependence of the intralayer transport time at the Fermi surface. 
We argue that, for certain systems, one should be
able to see Hall drag at intermediate magnetic fields and high
enough temperatures.

\section{Fallacious argument against Hall drag}
\label{sec:fallacious}

As mentioned previously, the quantity which is usually measured
experimentally is the transresistivity $\tensor\rho_{21}$, defined by 
\begin{equation}
\tensor\rho_{21}{\bf J}_1 = {\bf E}_2,
\end{equation}
with ${\bf J}_2 = 0$.
In the absence of a magnetic field in an isotropic system,
symmetry clearly dictates that ${\bf E}_2$ and ${\bf J}_1$ must be
parallel to each other. 
When a symmetry-breaking perpendicular $B$-field is imposed system,   
is it then possible to observe ``Hall drag'' voltage; i.e. a nonzero
off-diagonal element in $\tensor\rho_{21}$? 

From a macroscopic point of view, the following simple argument
seems to preclude the existence of Hall drag (barring quantum 
correlation effects which go beyond the Born approximation 
utilized in this paper).  In a drag experiment, the drive current
${\bf J}_1$ produces a parallel force ${\bf F}_{21}$ on the carriers 
in layer  2.
In steady state, the total net force on carriers in layer 2 must be zero.  
The additional forces acting on these carriers are the induced electric field, 
$e_2{\bf E}_2$ ($e_i$ is the charge in layer $i$), 
forces due to the imposed magnetic field ${\bf J}_2 \times {\bf B}$ 
and the lattice scattering.  
Since ${\bf J}_2 = 0$, both the Lorentz force and lattice scattering
are zero. Therefore, $e_2{\bf E}_2 = {\bf F}_{21}$, 
and since ${\bf F}_{21} |\!| {\bf J}_1$, ${\bf J}_1$ must also be 
parallel to ${\bf E}_2$.  That is, there should be no Hall drag.

This argument in specious on two counts.
First, ${\bf F}_{21}$ need not be parallel to ${\bf J}_1$.
The ${\bf F}_{21}$ depends on the exact nature of the distribution 
function $f_1({\bf k})$ of layer 1 in the presence of the driving
electric field.  When symmetry is broken by application of 
a magnetic field, $f_1({\bf k})$ may become skewed in a manner
which results in non-parallel ${\bf F}_{21}$ and ${\bf J}_1$.
Second, even though 
${\bf J}_2 = 0$, the carriers in layer 2 are not in equilibrium,
because they are continuously being acted upon by the drag force and 
induced electric field ${\bf E}_2$.  Therefore, since $f_2({\bf k})$ is not
necessarily equal to the equilibrium distribution function,
it is possible for the lattice to exert a net
force on the carriers in spite of the absence of a net current.

Thus, the presence or absence of measurable Hall drag in a Coulomb coupled
system depends crucially on the {\sl microscopic} details of the system. 
In particular, as we show below, Hall drag depends on the energy 
dependence of the {\sl intra}layer scattering mechanisms 
around the Fermi surface.  In this way, Hall drag measurements are 
distinct from the usual transport single layer Hall measurements, which are
generally quite insensitive to the details of the energy-dependence 
of the scattering mechanisms.  

\section{Formalism}

In this paper, we only treat cases where the $B$-field is small 
enough that Landau quantization is not significant (i.e., the
cyclotron frequency is much less than the inverse lifetime of the
electrons), 
and the interlayer interaction $W_{21}$ is weak so that one can 
work to the lowest non-vanishing (second) order in $W_{21}$;
i.e., in the Born approximation.
Given these assumptions, the semi-classical Boltzmann equation 
description is a valid description of the system.
We also assume that the carriers in an isotropic parabolic
band with effective mass $m_i^*$.  

The semi-classical theory gives the transconductivity 
$\tensor\sigma_{21}$ from which the transresistivity is obtained by 
\begin{equation}
\tensor\rho_{21} = -\tensor\rho_{22}\tensor\sigma_{21}\tensor\rho_{11}
\label{transresist}
\end{equation}
where $\tensor\rho_{ii}$ are the resistivity tensors of the individual
layers.
Following the formalism in Ref. \cite{flen95b},
generalized to include a $B$-field in the $z$-direction,
the transconductivity is given by
\begin{eqnarray}
\tensor\sigma_{21}({\bf B}) =  \frac{e_1 e_2}{8\pi k_B T}
\int \frac{d{\bf q}}{(2\pi)^2} \int_{0}^\infty
d\omega\ \frac{|W_{21}({\bf q},\omega)|^2}{\sinh^2(\hbar\omega/2k_B T)}
\nonumber\\
\ {\bf\Delta}_2({\bf q},\omega;-{\bf B})\ {\bf\Delta}_1({\bf
q},\omega;{\bf B}).
\label{sigma21}
\end{eqnarray}
The interlayer coupling $W_{21}$ is the screened Coulomb interaction
evaluated within the Thomas-Fermi approximation.

In the Kubo formalism, ${\bf\Delta}$ is given diagrammatically
by three Green functions arranged in a triangle\cite{kame95,flen95a}.   
In the Boltzmann formalism, ${\bf\Delta}$ is related
to the linear response in the distribution functions $f_i({\bf k})$ of 
the individual electron gases $i$ to a small uniform perturbing electric field.
Let ${\bf\Psi}_{{\bf B},i}({\bf k})$ be the quantity which describes the 
perturbation to $f_i({\bf k})$ which would result from the application
of a small electric field ${\bf E}_i$, 
in the presence of magnetic field {\bf B},
\begin{equation}
\delta f_i({\bf k}) \equiv f_i({\bf k}) - f^0_i({\bf k}) = 
f^0_i({\bf k})\,(1-f^0_i({\bf k}))\; e_i 
{\bf E}_i\cdot{\bf\Psi}_{{\bf B},i}({\bf k}),
\label{deltaf}
\end{equation}
where $f^0_i({\bf k})$ is the equilibrium Fermi-Dirac 
distribution function of layer $i$.
It can be shown that\cite{hu96b}
\begin{eqnarray}
{\bf\Delta }_{i}({\bf q},\omega;{\bf B}) &\equiv& 4\pi k_B T
\int \frac{d{\bf k}_i}{(2\pi)^2}\
[{\bf\Psi}_{{\bf B},i}({\bf k}_i+{\bf q}) -
{\bf\Psi}_{{\bf B},i}({\bf k}_i)]\nonumber\\ 
&&[f_i^0({\bf k}_i) - f_i^0({\bf k}_i+{\bf q})]\ 
\delta(\varepsilon_{\bf k_i}-
\varepsilon_{\bf k_i + q} -\hbar\omega),
\label{Delta}
\end{eqnarray}
Furthermore, the single layer conductivities are also related 
to ${\bf\Psi}$ by\cite{hu96b}
\begin{equation}
\tensor\sigma_{ii}({\bf B}) = 2 e_i^2 k_B T \int
\frac{d{\bf k}}{(2\pi)^2}\
\left(-{f^0_i}^\prime (\varepsilon)\right)\,
{\bf v}_{{\bf k},i}\, \Psi_{{\bf B},i}({\bf k})
\label{sigmaii}
\end{equation}
and $\tensor\rho_{ii}$ can be obtained by inverting $\tensor\sigma_{ii}$.

\subsection{$\protect\Psi$ in a $B$-field} 

We assume the intralayer scattering of the system dominated by 
elastic (e.g., impurity) and 
quasi-elastic (e.g., acoustic phonon) scattering, as is the case for
GaAs under 40 K.  Under these circumstances, the scattering can be 
described by an energy-dependent transport 
time $\tau(\varepsilon)$,
whose exact functional form of course depends on the particular system
being studied. Then, ${\bf\Psi}_{{\bf B},i}$ is\cite{askerov}
\begin{eqnarray}
{\bf\Psi_{{\bf B},i}(k)} &=& \frac{\tau_i(\varepsilon_k)}{k_B T}
\frac{{\bf v}_{{\bf k},i} - {\bf v}_{{\bf k},i}\times \hat{\bf z}  \,
(\omega_{c,i}\tau_{i}(\varepsilon_k))}
{1 + (\omega_{c,i}\tau_{i}(\varepsilon_k))^2},\nonumber\\
&=&\frac{\tau_i(\varepsilon_k) v_k}{k_B T \sqrt{1 +
(\omega_c\tau_i(\varepsilon_k))^2}}\ 
\hat{\bf a}_i(\varepsilon_k)
\label{Psi}
\end{eqnarray}
where ${\bf v}_{\bf k}$ is the velocity, $\omega_{c,i} = e_iB/m^*_i$ 
is the cyclotron frequency 
and $\hat{\bf a}$ is a unit vector rotated at an angle 
$-\tan^{-1}(\omega_{c,i}\tau(\varepsilon_k))$ from ${\bf k}$.
This is shown schematically in Fig.\ 2.

\subsection{Energy-independent $\protect\tau_i$}

In the case when the $\tau_1$ is energy-independent,
$\hat{\bf a}_1$ is a constant.  
Then, Eq.\ (\ref{Psi}) shows that $\delta f_1$ is inversion 
symmetric with respect to $\hat{\bf a}_1$, which implies that 
the current ${\bf J}_1$ is parallel to ${\bf a}_1$ and,   
from the Born approximation expression for the force transferred 
from layer 1 to 2\cite{jauh93}, that ${\bf F}_{21}$ is parallel to 
${\bf J}_1$.  Furthermore, the net lattice force on carriers in layer 2 
for energy-independent $\tau_2$ is simply proportional to ${\bf J}_2$, 
and hence is zero in a transresistivity experiment.  
Thus, in this special case, the specious arguments given in 
Sec.\ \ref{sec:fallacious} actually hold, and there is no Hall drag. 

\subsection{Energy-dependent $\protect\tau_\protect{i}$}

However, the sophistry of the argument is exposed once 
$\tau_i$ is energy-dependent.  The energy dependence of 
$\hat{\bf a}_1$ then results in a $\delta f_1$ which is no 
longer inversion symmetric with respect to the axis of ${\bf J}_1$, 
and consequently ${\bf F}_{21}$ is not necessarily parallel to ${\bf J}_1$.  
Furthermore, the lattice can exert a non-zero force on the carriers in
layer 2 (despite ${\bf J}_2$ being zero) which is non-parallel 
to ${\bf F}_{21}$.  Thus, in principle Hall drag can exist. 

\section{Magnitude of Hall drag: small $T$ expansion}

Merely giving an existence argument is insufficient;
one would like to know if the effect is experimentally observable.  
We address this point in this section, by calculating the low temperature 
behavior of $\tensor\rho_{21}^{xy}$. 

We first linearise the energy-dependence of the transport time 
about the Fermi energy $\varepsilon_{F}$,
\begin{equation}
\tau(\varepsilon) = \tau_0 \left( 1+ s\frac{\xi}{\varepsilon_F}\right),  
\end{equation}
$\xi = \varepsilon-\varepsilon_F.$
We also find it convenient to write the conductivity and resistivity tensors
in terms of a product of a scalar and rotation
matrix 
\begin{equation}
\hat R(\theta) = \left(\matrix{\cos\theta& -\sin\theta\cr
			       \sin\theta& \cos\theta}\right)
\end{equation}
which rotates vectors clockwise by angle $\theta$.
The magnitude and rotation angle of $\tensor\rho$ ($\tensor\sigma$)
are denoted by $\varrho$ ($\varsigma$) and $\theta$ ($\phi$),
respectively; i.e., 
\begin{eqnarray}
\tensor\rho_{ij}(B,T) &=& 
\varrho_{ij}(B,T)\; \hat R\Bigl(\theta_{ij}(B,T)\Bigr);\\
\tensor\sigma_{21}(B,T) &=& 
\varsigma_{ij}(B,T)\; \hat R\Bigl(\phi_{ij}(B,T)\Bigr).
\label{rhosigma}
\end{eqnarray}
Therefore, from Eq. (\ref{transresist}),
\begin{eqnarray}
\varrho_{21} &=& -\varrho_{22}\,\varsigma_{21}\,
\varrho_{11};\label{varrho21} \\
\theta_{21} &=& \theta_{22}+\phi_{21}+\theta_{11}.
\label{rho_21} \label{theta21}
\end{eqnarray}
We denote the $n$-th coefficient of an expansion in powers of $T$ 
of quantity $A$ as $A^{(n)}$; e.g., 
\begin{eqnarray}
\theta_{ij}(B,T) &=& \sum_{n=0}^\infty 
\theta_{ij}^{(n)}(B)\; T^n.
\end{eqnarray}

\subsection{The $T\rightarrow 0$ limit}

First, let us briefly review some results of Coulomb drag in the
absence of a $B$-field. 
At $B=0$, all the rotation angles in an isotropic system 
are clearly zero by symmetry.  
Due to phase space considerations, in the low temperature 
limit\cite{nodiff}
$\sigma_{21}(B=0,T) = \varsigma_{21}^{(2)}(0) T^2 + O(T^3)$,
as in $e$--$e$ scattering in a single two-dimensional layer
\cite{2dscat}.  Since $\varrho_{ii}^{(0)} (B=0) = m^*_i/(n_ie_i^2\tau_{0,i})$ 
($n_i$ is the density), Eq.\ (\ref{transresist}) implies that 
$\rho_{21}(B=0,T)$ also has a quadratic temperature dependence.  
The leading order coefficients 
$\varrho_{21}^{(2)}(B=0)$ and $\varsigma_{21}^{(0)}(B=0)$
are given elsewhere\cite{jauh93,flen95b}.

In the presence of a magnetic field, 
Eqs. (\ref{sigma21}) and (\ref{sigmaii}),
together with Eqs.\ (\ref{Delta}) and (\ref{Psi}),
yield to lowest non-vanishing order in $T$ 
\begin{eqnarray}
\varrho_{ii}^{(0)}(B) &=& \varrho_{ii}^{(0)}(B=0) 
\left( 1 + \Omega_{c,i}^2\right)^{1/2}\label{varrhoii}\\
\varsigma_{21}^{(2)}(B) &=& 
\frac{\varsigma_{21}^{(2)}(B=0)}{[(1+\Omega_{c,1}^2)
(1+\Omega_{c,2}^2)]^{1/2}}
\\
\theta_{ii}^{(0)}(B) &=& \tan^{-1}(\Omega_{c,i});\\
\phi_{21}^{(0)}(B) &=& -\tan^{-1}(\Omega_{c,1}) - \tan^{-1}(\Omega_{c,2}),
\label{theta0}
\end{eqnarray}
where $\Omega_{c,i} = \omega_{c,i}\tau_{0,i}$.
From Eqs. (\ref{varrho21}), (\ref{theta21}) and 
(\ref{varrhoii}) -- (\ref{theta0}), we obtain
$\theta_{\rho,21}^{(0)}(B) = 0$ and 
\begin{eqnarray}
\varrho_{21}^{(2)}(B) &=& \varrho_{21}^{(2)}(B=0).
\end{eqnarray}
This means that $\tensor\rho_{21}(B,T\rightarrow 0)$ 
is {\sl independent} of the magnetic field, and 
hence the ratio of the Hall to normal drag coefficients in this limit is
\begin{equation}
\lim_{T\rightarrow 0}\ \frac{\rho^{xy}_{21}(B,T)}{\rho^{xx}_{21}(B,T)}= 0.
\end{equation}

The fact that the Hall drag disappears {\sl faster} than normal drag 
as $T\rightarrow 0$ is a consequence of the lack of symmetry-breaking 
in the distribution function in this limit. 
From Eq. (\ref{deltaf}), one sees that 
$\delta f_i({\bf k})$ is significant only around 
an energy of order of a few $k_B T$ about the
Fermi energy.  If the scattering time $\tau_i(\varepsilon)$ 
does not change significantly within this energy range, then
the symmetry breaking in $\delta f_i$ will be small, and consequently
so will the Hall drag effect. 
One needs to go to larger temperature to see a measurable Hall drag 
signal.

To obtain the first non-vanishing term in the $T$-expansion of
$\rho_{21}^{xy}$, we expand the rotation angles $\theta_{ii}$
and $\phi_{21}$ in powers of $T$.
This is achieved by expanding $\Psi$ in powers of 
$\xi$, and using this expansion in Eq. (\ref{sigmaii}).
Inverting $\tensor\sigma_{ii}$, we find $\theta_{ii}^{(1)} = 0$ 
and 
\begin{eqnarray}
\theta_{ii}^{(2)} &=& \frac{\pi^2 s_i \Omega_{c,i} ( 1 + s_i + 
\Omega_{c,i}^2 - s_i\Omega_{c,i}^2)}{3 (1 + \Omega_{c,i}^2)^2}\;
\frac{k_B^2}{\varepsilon_{F,i}^2}.
\label{thetaii_2}
\end{eqnarray}
The angle $\theta_{ii}$ increases with increasing $s$ 
(for $\Omega_{c,i}^2 < 1$) because the particles with larger velocities,
which contribute more to the overall conductivity, have a larger 
deflection with respect to ${\bf E}_i$ (see Eq. (\ref{Psi})).

At this point to simplify the algebra (which otherwise would be daunting), 
it is henceforth assumed that both the layers are identical, 
and therefore the layer indices for all the parameters shall be dropped.
We also assume that the well widths $L$ are zero, and the 
inter-well spacing $d$, the Fermi wavevector
$k_F$ and the the Thomas-Fermi screening length $q_{\rm TF}$
satisfy the conditions $(k_F d)^{-1} \ll 1$ and 
$(q_{\rm TF} d)^{-1} \ll 1$.   Hence, the results presented here are 
only valid to lowest order in these quantities.

Expanding ${\bf\Delta}$ in powers of $\omega$ in Eq. (\ref{sigma21}) 
yields $\phi_{21}^{(1)}(B) = 0$ and
\begin{eqnarray}
\phi_{21}^{(2)}(B) &=& 
\frac{\pi^2 s \Omega_c (1 - 2 s + \Omega_c^2 + 2 s \Omega_c^2)}
{3 (1+\Omega_c^2)^2}\; \frac{k_B^2}{\varepsilon_F^2}
\label{phi21_2}
\end{eqnarray}
From Eqs. (\ref{theta21}), (\ref{thetaii_2}) and 
(\ref{phi21_2}), the small $T$ rotation angle of $\tensor\rho_{21}$ is 
\begin{equation}
\theta_{21}(T) \approx      
\frac{\pi^2 s \Omega_c}{(1 + \Omega_c^2)}\;\frac{(k_B T)^2}{\varepsilon_F^2}.
\end{equation}
For small angles, $\sin\theta \approx \theta$, and hence that
the ratio of the Hall to longitudinal transresistivities is 
coefficient is
\begin{equation}
\frac{\rho_{21}^{xy}(B,T)}{\rho_{21}^{xx}(B,T)}
= \frac{\pi^2 s \Omega_c}{(1 + \Omega_c^2)}\;
\frac{(k_B T)^2}{\varepsilon_F^2} + O(T^3)
\label{ratio}
\end{equation}
Since $\rho_{21}^{xx} \propto T^2$, this shows that 
$\rho_{21}^{xy}\propto T^4$.

For positive $s$ and like charges in layers 1 and 2, 
both $\theta_{11}$ and $\theta_{21}$ have the same sign. 
Since $\varrho_{ii}^{(0)}$ and $\varsigma_{ii}^{(2)}$ (for like
charges) are positive, the negative sign in Eq.\ (\ref{varrho21}) 
means that the Hall fields in the driving and drag layer are in 
{\sl opposite} directions.  From an experimental point of view,
this is favourable because a Hall drag signal cannot be mistaken 
for a leakage voltage from the driving layer\cite{gram_priv}.

\section{Discussion}

Since the magnitude of $\rho_{21}^{xy}$ is proportional to $s$,
one would like to have a large value of $s$ to 
obtain an experimentally clear signal.  
Herein lies a problem.  Generally in the modulation doped samples
currently used in drag experiments, the remote dopants are placed far 
away in order to obtain high mobilities in the quantum wells. 
This means that the {\sl intra}layer $e$--$e$ scattering 
is much stronger than either the impurity scattering or 
acoustic phonon scattering, and hence even when the system
is driven by external forces, the $f({\bf k})$ tends to
relax towards a drifted Fermi-Dirac distribution.  
A drifted Fermi-Dirac distribution function 
is equivalent to a ${\bf\Psi}$ in Eq. (\ref{Psi}) with 
a constant $\tau$; i.e., with $s=0$.  Therefore, when intralayer
$e$--$e$ scattering dominates, Hall drag will be difficult
to measure. 

To get a measurable Hall drag signal, one needs to increase $s$.  
This can be done by putting the charged dopants close to the quantum
wells.  As shown in Ref.\ \cite{hu96a}, when the dopants are placed
on the order of $150\;{\rm\AA}$ from the side of a GaAs well doped
to $1.5\times 10^{11}\,{\rm cm}^{-2}$, one can achieve an $s$ 
on the order of 0.4.  The factor $\Omega_{c}/(1+\Omega_c^2)$
has a maximum of 1/2 at $\Omega_c = 1$\cite{validity},
and therefore the prefactor in Eq. (\ref{ratio}), 
$\pi^2 s \Omega_c/(1+\Omega_c^2)$ can be made larger than one, which 
should facilitate measurement of Hall drag.

To conclude, we have shown that it is possible to measure Hall drag 
in Coulomb coupled quantum wells.  The Hall drag coefficient
$\rho_{21}^{xy}$ goes as $T^4$, and it probes the 
$\varepsilon$ dependence of the transport time $\tau(\varepsilon)$ in the
vicinity of the Fermi energy.  Note that the single layer Hall coefficient 
also depends to some extent on the $\varepsilon$ dependence of 
$\tau(\varepsilon)$ through the Hall coefficient\cite{seeger} 
$r_H = \langle \tau^2(\varepsilon) \rangle
/\langle\tau(\varepsilon)\rangle^2$ (where $\langle\cdots\rangle$
denotes thermal averaging).  However, the $\varepsilon$-dependence 
in $\tau(\varepsilon)$ gives a correction factor to $r_H$, 
whereas it affects Hall drag to leading order, and therefore
Hall drag is a much more sensitive probe of $\tau(\varepsilon)$.    

\section{Acknowledgement}

We thank Martin Christian B\o nsager and Karsten Flensberg
for useful discussions.

%\bibliographystyle{pscript}
%\bibliography{halldrag}

\begin{figure}
\epsfxsize=9.3cm
\hspace*{3.0cm}
\epsfbox{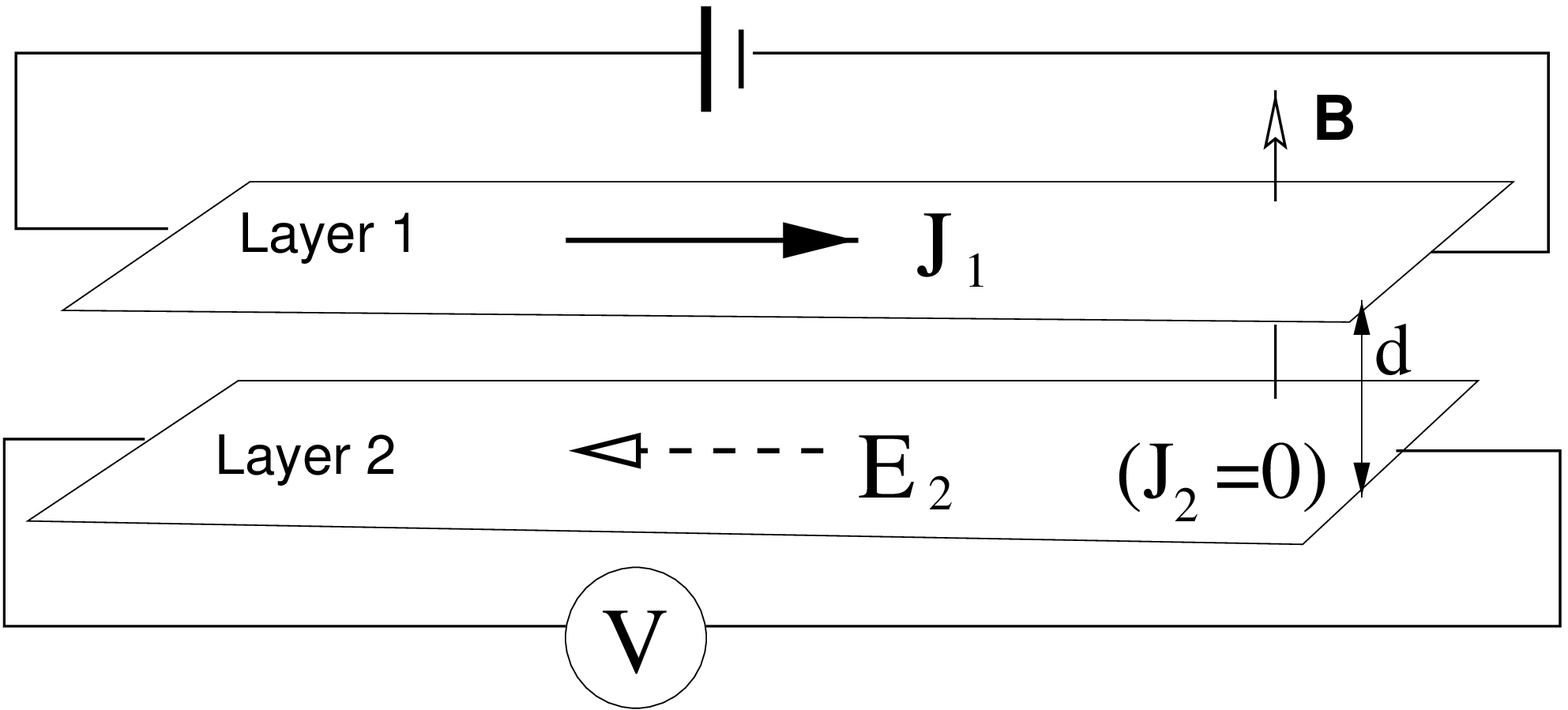}
\bigskip\bigskip
\caption{Schematic diagram of drag experiment.  Two independently
contacted two-dimensional electron gases are placed close together.
A current ${\protect\bf J}_1$ is driven through layer 1, and layer 2
is connected to a voltmeter (so that ${\protect\bf J}_2=0$).  
The interlayer $e$--$e$ interactions cause a 
drag response electric field ${\protect\bf E}_2$ in layer 2. 
If a magnetic field ${\bf B}$ perpendicular
to both layers is applied, can ${\protect\bf E}_2$ have a component 
perpendicular to ${\bf J}_1$?  We show that this should be possible 
for high enough temperatures and large enough energy dependence
of intralayer transport times.}
\bigskip\bigskip\bigskip\bigskip\bigskip

\epsfxsize=7.0cm
\hspace*{4.0cm}
\epsfbox{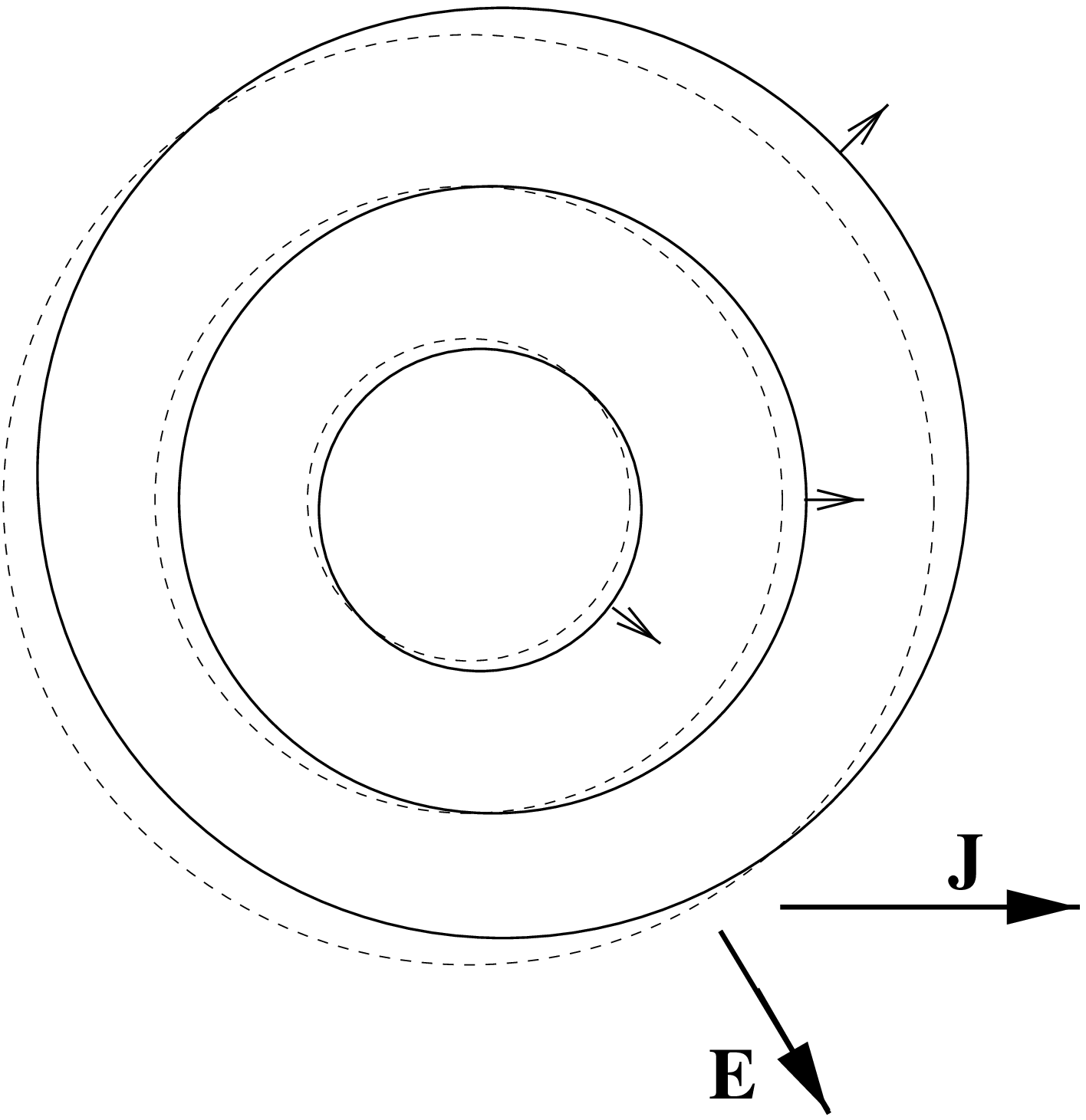}
\bigskip\bigskip
\caption{Schematic contour plot of distribution function of carriers 
in the ${\bf k}$-plane, in an applied magnetic field for 
non-constant $\tau(\varepsilon)$.  
The dotted and solid lines are for carriers in equilibrium and  
in an electric field ${\bf E}$, respectively.
When ${\bf E}$ is applied, the contours shift in different directions,
as shown by the small arrows, due to the variation $\tau(\varepsilon)$ 
[see Eq.  (\ref{Psi})].  The resultant net current is ${\bf J}$.}
\label{fig2}
\bigskip
\end{figure}

\end{document}